# Charged Obstacles Augment Electrokinetic Energy Conversion Efficiency

Tanoy Kahali, Antarip Poddar, Jayabrata Dhar and Suman Chakraborty

Department of Mechanical Engineering, Indian Institute of Technology Kharagpur, Kharagpur, India-721302

**ABSTRACT**

In the present study, we delineate the effect of introducing flow obstructions on streaming potential and energy conversion efficiency in a narrow fluidic confinement taking into consideration the wall hydrodynamic slip, finite ionic size, and local permittivity variation effects. We consider two types of the geometric pattern of flow obstacle between which regular pattern turns out to be more effective. It is observed that, implementing structured flow obstacles of radii greater than 10% of channel height and charge density ratio (charge density of flow obstacle surface to that of channel wall) greater than or equal to 9 renders significant enhancement of power generation efficiency. On the other hand, it is effective to operate at a charge density ratio~1 for obstacles radii less than or equal to 10% of channel height. An effective normalized pitch length of magnitude 0.6 or above has to be maintained in order to obtain optimum energy conversion efficiency. We found that implementation of charged flow obstacle causes a significant enhancement of energy conversion efficiency (~18% with regular pattern) comparing to planar slit channel (~4%) without considering wall hydrodynamic slip and steric effects. Introducing finite ionic size and wall hydrodynamic slip effect, pertinent to flows in narrow confinement, leads to further enhancement in electrokinetic energy conversion efficiency.

## 1. INTRODUCTION

Modern micro fabrication technologies have brought about the era of 'Micro total analysis system' (μ-TAS) and 'Lab on a chip' devices providing a more convenient platform for rapid testing, study and analysis of sample along with efficient multiple functionalities towards flow actuation, particle separation, micro mixing, energy generation, biomedical engineering, electronic chip cooling [1–8] to name a few. Shrinking the length scale down to a few hundred μm or lower leads a significant alteration of transport phenomena. One of the central aspects of micro fluidic transport is electro-kinetics [9] which deals with the dynamics of fluid flow in narrow conduits in presence of an electrical double layer (EDL). EDL plays a pivotal role in the establishment of all electro-kinetic phenomena and helps in flow manipulation using the electric field. In recent times, electro-kinetic effects especially electro-osmotic pumping [10,11] and electrokinetic energy conversion [12–21] have been widely investigated in micro and nanofluidic devices. Due to the presence of excess counter-ions (oppositely charged ions with respect to the surface charge) in bulk (diffuse layer), an externally imposed pressure gradient causes advection of these ions towards the downstream which in turn induces a net potential difference across two terminals of the channel widely known as streaming potential [22–24] and the current generated is known as streaming current. This has to balance the conduction current (neglecting Stern layer conductance) in order to satisfy the overall electro-neutrality condition in absence of any externally applied electrical load. Induced streaming field gives rise to a back electro-osmotic flow which in turn opposes the forward pressure driven flow causing reduction of flow rate. However, on a beneficial note, the streaming potential causes a flow of net current due to the presence of an external load resistor attached across the tapped ends of a channel and thus provides means for the conversion of the hydraulic to electrical power.[13,15,20,25,26] Such EKEC has been the topic of interest, both from theoretical and experimental perspectives[27]. Different theoretical investigations [13,15–17] reported significant augmentation of streaming potential and energy conversion efficiency considering wall slip in planar slit channel where hydrodynamic slip length ranges from a few nm [15,28–32] up to 33mm.[33,34]. The role of finite ionic size consideration at high surface charge density was also found to increase the streaming potential [17,35] and EKEC.[25]

In line with the above studies, electro-osmotic flow (EOF) in porous media has been widely investigated experimentally and numerically as well by many research groups for both Newtonian and non-Newtonian fluids. [36–39] The flow rate has been found to increase with an increase in the size of solid obstructions [10,38,40–45] and porosity.[38,41,43] Besides, the effect of pitch length ( center to center distance between two consecutive obstacles) on EOF was analyzed extensively in past decades. However, most of these numerical modelings are based on the simplified hypothesis of the governing equations(charge distribution on the surface of the flow obstructions is disregarded in both the governing equations for internal potential distribution and hydro-dynamics)[46]. In their recent work, Li and coworkers[37] reported a maximum achievable EOF velocity corresponding to a gap between two consecutive obstacles equal to five times the Debye length and found to decrease with any deviation from that value. For a very narrow spacing between two obstacles, their respective EDL overlaps leading to a prominent degradation of the flow speed. Later, Scales and Tait [47] investigated the influence of such flow obstructions on the electroosmotic flow rate in micro channel introducing circular flow obstructions of different size and porosity. They reported an augmentation in flow rate for a channel dimension greater than a critical value. Yang et al.[32] proposed a theoretical model of a battery using water as working fluid passing through a porous glass filter. They concluded that higher salt concentration leads to an enhancement of streaming current.

From the above discussion, it is observed that most of the works, especially on EKEC, are concentrated on planar channel or capillaries. Though, a significant effort has been made in studying the streaming potential and EKEC in presence of nano-porous membrane, the consequence of structured charged obstacles on the same remains unaddressed. In the present study, we investigate the effect of structured flow obstacle size, pitch-length, the ratio of surface charge density of the pore to channel wall, pore arrangement (regular/zig-zag) on streaming potential and energy conversion efficiency. Besides considering a modified version of the Poisson-Nernst-Plank (mPNP) equation [48] to account for the finite ionic size, the present model adopts a realistic approach by considering wall-hydrodynamic slip and interfacial permittivity variation.

**2. PROBLEM FORMULATION**:

We consider the pressure driven flow of an electrolyte solution along the positive *x* direction through a narrow fluidic channel. Stream wise and cross stream wise coordinates are denoted by *x* and *y* respectively. Coordinate axis is attached to the lower plate of the channel. Due to the interaction of dielectric walls of both channel and the solid obstacle surface with the polar fluid, there is a migration of ions across their interface resulting in a charged wall substrate.

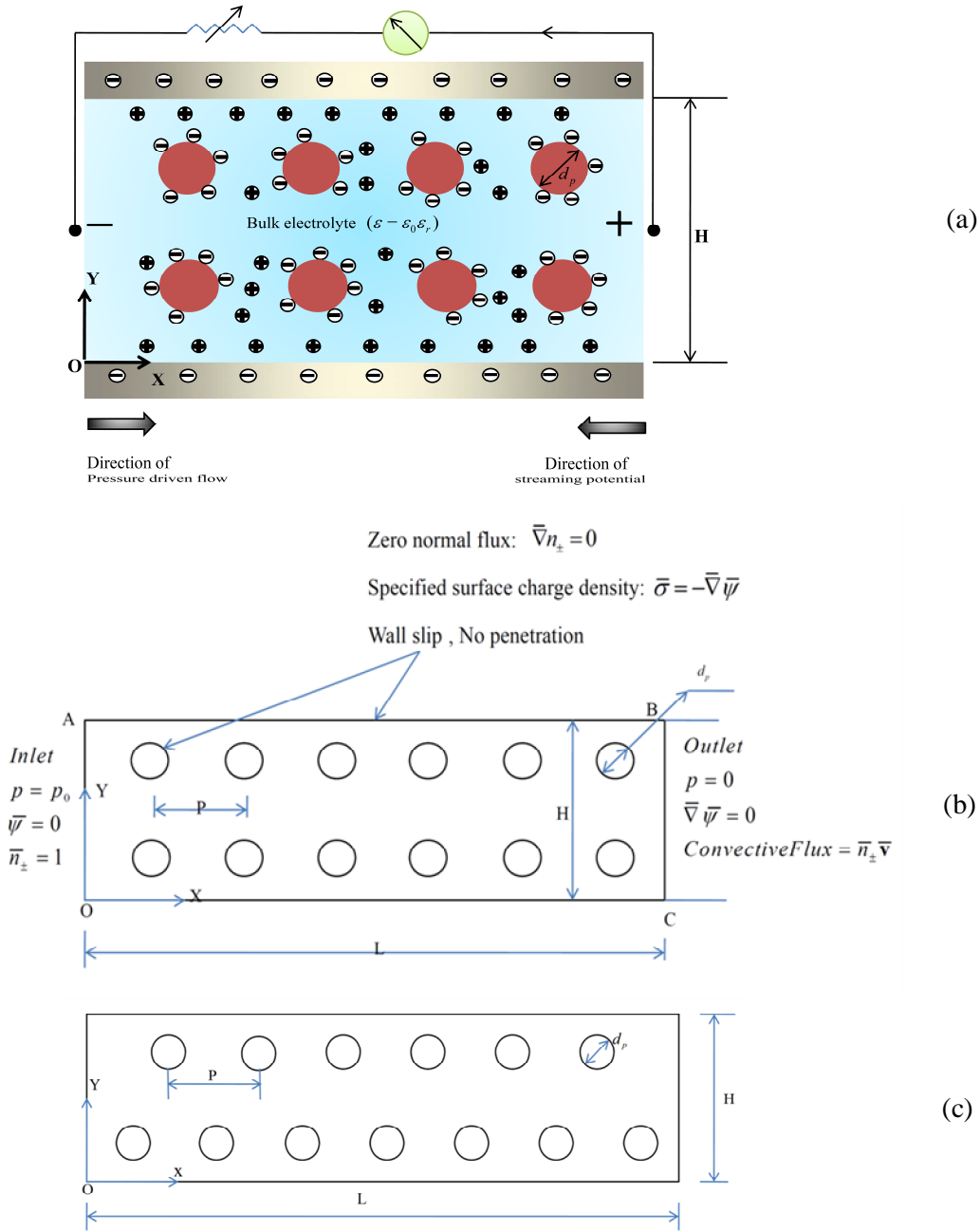

**Fig.1(a)** Flow of electrolyte solution is taking place through a narrow fluidic channel of height H, having solid cylindrical flow obstructions of diameter $d_p = 2R_p$ arranged in a regular pattern in form of a $N \times M$ matrix.(where $N$ and $M$ denotes number of structured flow obstacle in a single row and number of rows respectively). Solid surface of the channel and the solid obstacles are assumed to adsorb negative charge. Direction of pressure driven flow and induced potential is shown in the above figure. Direction of streaming current is same as the direction of pressure driven flow. **Fig**.-**(b)** & **(c)** delineates configuration of flow obstructions used in the present simulation where **(b)** denotes regular (geometry-1) and **(c)** denotes zigzag pattern (geometry-2) of flow obstacles.

### 2.1.1 Hydrodynamics

Surface electro-kinetics plays a pivotal role in describing the flow physics for flow through narrow confinement. The electrical effect is manifested in the momentum balance equation by implementing a body force term consisting of the volumetric charge density of the ions and the electrical field strength. By assuming steady, incompressible flow at very low Reynolds number, thereby eliminating inertial term from the equation, Navier-Stokes equation simply boils down to Stokes equation

$$0 = -\nabla P + \mu \nabla^2 \mathbf{v} - \rho_f \nabla \psi \qquad (1)$$

$$\nabla \cdot \mathbf{v} = 0 \qquad (2)$$

where $\mathbf{v} = (u, v)$ denotes the velocity vector having x and y velocity components as $u$ and $v$ respectively; $\mu$ is the fluid viscosity; $\rho_f$ is the volumetric charge density of the ions present in the system and $\psi$ is the electrical potential.

The last term of equation (1) is the body force term originated due to electrical interaction. It is noteworthy that the electrical field $(-\nabla \psi)$ appearing in the body force term is an induced one due to the advection and accumulation of ions in the downstream caused by the externally applied pressure gradient. To estimate the body force due to induced electric potential, the electrostatic potential distribution within the EDL and the distribution of ionic species are to be determined.

### 2.1.2 Charge distribution within EDL

Relation between volumetric charge density and electrical potential is manifested through Poisson equation which can be expressed as

$$\nabla \cdot (\varepsilon_r \nabla \psi) = -\frac{\rho_f}{\varepsilon_0} \qquad (3)$$

Volumetric charge density can be written as $\rho_f = \sum z_i e n_i$, where $z_i$ is the valence of the $i$-th ionic species; $e$ is the protonic charge and $n_i$ is the ionic number concentration of $i$-th species. Assuming a 1:1 symmetric electrolyte dissociation, the volumetric charge density can be written as $\rho_f = ze(n_+ - n_-)$, where $z$ denotes the valence of the ion $(z_+ = z, z_- = -z)$; $n_+$ and $n_-$ denotes the ionic number density for positive and negatively charged species; $\varepsilon_0$ and $\varepsilon_r$ denotes permittivity of vacuum and relative permittivity of the electrolyte respectively.

### 2.1.3 Transport of Ionic species

*Modification in PNP equation incorporating steric effect*

In the case of high surface charge density, Poisson-Boltzmann equation has severe shortcomings because it considers the ions as point mass (neglects its finite size) and neglects ion-ion interactions. Besides, macroscopic advection and diffusion of ionic species are neglected in the assumption of well celebrated Boltzmann distribution. Hence, in the present model, ionic transport is governed by very fundamental Poisson-Nernst-Plank (PNP) equation with a correction term due to steric effect [48]

$$\frac{\partial n_\pm}{\partial t} = D\nabla^2 n_\pm + \frac{Dez_\pm}{k_B T}\nabla \cdot (n_\pm \nabla \psi) - \nabla \cdot (n_\pm \mathbf{u}) + a^3 D \nabla \cdot \left( \frac{n_\pm \nabla (n_+ + n_-)}{1 - a^3(n_+ + n_-)} \right) \qquad (4)$$

where, D is the diffusion coefficient of ionic species. Here, we assume $D_+ = D_- = D$. Ionic mobility is expressed as $b_+ = b_- = b = D/k_B T$; $T$ denotes temperature in Kelvin and $k_B$ is the Boltzmann constant. The last term in equation (4) is the correction term due to the incorporation of finite size effect; $a$ denotes typical spacing between two densely packed ions. The steric factor $v$ is related to $a$ as $v = 2a^3 n_\infty$, where $n_\infty$ denotes bulk ionic concentration.

### 2.1.4 Effect of local permittivity variation

Two main reasons causing permittivity variation in the interfacial region are a) depletion of water molecules adjacent to the charged surface and (b) increase in the orientation ordering of water dipoles in presence of high charge density. In the present work, a modified relative permittivity expression [49] is used throughout taking into account the spatially oriented dipoles of water molecules (polarization) and finite ionic size. It is assumed that in the bulk, the position in a 'lattice' can be occupied either by counter-ions, co-ions or water molecules. Hence, the number density of lattice site $(n_s)$ in the bulk is a constant and can be expressed as $n_s = 2n_\infty + n_{0,w}$, where $n_{0,w}$ is the number density of water molecules and $n_\infty$ is for both counter and co-ions in the bulk. Number density variation of counter and co-ions in the bulk can be expressed as follows:

$$n_+ = n_- = n_s \left( \frac{n_\infty}{2n_\infty + n_{0,w}} \right) \tag{5}$$

$$n_w = n_s \left( \frac{n_{0,w}}{2n_\infty + n_{0,w}} \right) \tag{6}$$

The relative permittivity can be given as follows ([35,49]):

$$\varepsilon_r = 1 + n_s n_{0,w} \left( \frac{p_0 F(f)}{R|\nabla \psi|} \right) \tag{7}$$

where $p_0$ is the magnitude of water dipole moment; $f = -p_0 |\nabla \psi|/k_b T$; $R$ is defined as $R = 2n_\infty \cosh(ze\psi/k_b T) + n_{0,w} \sinh(f)/(f)$; $F$ is a function given by $F(f) = L(f)(\sinh(f))/(f)$; $L(f) = [\coth(f) - 1/(f)]$ is the Langevin function.

**2.2 Non-dimensionalization**

In an aim to non-dimensionalize the governing equations (1-4), following nondimensional parameters are introduced:

$$\overline{\mathbf{v}} = \frac{\mathbf{v}}{u_{ref}}, \overline{x}_i = \frac{x_i}{L_{ref}}, \overline{\psi} = \frac{ze\psi}{k_b T}, \overline{n}_\pm = \frac{n_\pm}{n_\infty}, \overline{\nabla} = \nabla L_{ref},$$

$$\overline{\lambda} = \frac{\lambda}{L_{ref}}, \overline{\sigma} = \frac{zeL_{ref}\sigma}{k_b T \varepsilon_0 \varepsilon_r}, L_s = \beta/L_{ref}, p = L_{pitch}/L_{ref}$$

where $u_{ref} = \frac{L_{ref}^2}{\mu}(-\nabla P)$, $L_{ref} = H$, $x_i = x, y$ (for $i = 1 \& 2$ respectively); $L_{ref}$ is the reference length scale of the problem which is considered as the height of the channel in the present study. Dimensional and normalized hydrodynamic slip lengths are denoted by $\beta$ and $L_s$ respectively. $L_{pitch}$ and $p$ denotes dimensional and normalized pitch length respectively. The characteristic EDL thickness is denoted as $\lambda = \sqrt{\left(\varepsilon_0 k_b T / n_\infty z^2 e^2\right)}$ for a symmetric $(z:z)$ binary electrolyte and $\kappa = \lambda^{-1}$ is the inverse Debye length.

Hydrodynamic parameters like $\mu, \rho$ and diffusion coefficients $\left(D_\pm\right)$ are considered as constant. By substituting the above-mentioned non dimensional parameters, the equations (1-4) boil down to the following non dimensional form

$$0 = 1 + \overline{\nabla}^2 \overline{\mathbf{v}} + u_r \left[\overline{\nabla}.(\varepsilon_r \overline{\nabla} \overline{\psi})\right] \overline{\nabla} \overline{\psi} \tag{a}$$

$$\overline{\nabla}.\overline{\mathbf{v}} = 0 \tag{b}$$

$$\overline{\nabla}.\left(\varepsilon_r \overline{\nabla} \overline{\psi}\right) = -\left(\frac{\overline{n}_+ - \overline{n}_-}{\overline{\lambda}^2}\right) \tag{c} \quad (8)$$

$$\overline{\nabla}.\left(\overline{n}_\pm \overline{\mathbf{v}}\right) = \frac{1}{Pe}\overline{\nabla}.\left(\overline{\nabla}\overline{n}_\pm\right) + \left(\frac{z_i}{z}\right)\left(\frac{1}{Pe}\right)\overline{\nabla}.\left(\overline{n}_\pm \overline{\nabla} \overline{\psi}\right) + \left(\frac{0.5\nu}{Pe}\right)\overline{\nabla}.\left[\frac{\overline{n}_\pm \overline{\nabla}\left(\overline{n}_+ + \overline{n}_-\right)}{1 - 0.5\nu\left(\overline{n}_+ + \overline{n}_-\right)}\right] \tag{d}$$

We are interested in steady state response of the system. Here, $Pe = \left(u_{ref} L_{ref}/D\right)$ denotes ionic Péclet number; $u_r = \left(u_e/u_{ref}\right)$ denotes the relative strength between purely pressure driven flow velocity scale and reverse electro-osmotic velocity scale given as $u_{ref} = \left(L_{ref}^2/\mu\right)(-\nabla P)$ and $u_e = \left(\varepsilon_0 k_b^2 T^2 / z^2 e^2 \mu H\right)$, respectively.

### 2.3 Boundary conditions

A finite hydrodynamic slip length $\left(L_s\right)$ is considered at the walls of the channel as well as the obstacle surfaces (Fig.-1 (b)). Flow actuation is accomplished by employing a pressure gradient between the two terminals (AO and BC) of the channel. Surface charge density $\left(\overline{\sigma} = -\overline{\nabla}\overline{\psi}\right)$ is specified at channel walls and obstruction surfaces along with a zero potential gradient $\left(\overline{\nabla}\overline{\psi} = 0\right)$ at the exit (boundary BC). Also, the potential is set to zero at inlet boundary AO. Electrolyte solution having bulk concentration $\left(\overline{n}_\pm = 1\right)$ enters the channel (at AO). At the exit boundary (BC), a normal outwardly directed convective $\left(\overline{\mathbf{J}}_{exit} = \overline{n}_\pm \overline{\mathbf{v}}\right)$ flux boundary condition is employed, depicting the absence of any concentration gradient normal to the exit boundary.

## 2.4 Electro neutrality: induced streaming potential field and streaming current

Advection of ions to the downstream direction causes the flow of a current known as advection current or streaming current, $(I_{stream})$ which is balanced by conduction current $(I_{cond})$ flowing through bulk (neglecting Stern layer conductance). This satisfies the overall electro neutrality of the system given as: $I_{ionic} = (I_{stream} + I_{cond}) = 0$.

Streaming current flowing through the channel can be given as:

$$I_{stream} = ze \iint_A (n_+ - n_-) u \, dA \tag{9}$$

where $A$ is the channel cross sectional area. Considering width $w \gg H$, the expression of streaming current boils down to

$$I_{stream} = wze \int_{y=0}^{H} (n_+ - n_-) u \, dy \tag{10}$$

In most of the experimental devices, the streaming potential is calculated by measuring the potential difference between the reservoirs connecting the channel. However, the numerical modelling of transient streaming potential in a narrow cylindrical confinement with reservoir performed by Mansouri and co-workers [50] shows that, in cases where reservoir dimension is much larger (almost five times) than the channel radius, the flow field is hardly affected by the entry and exit effects of the channel.

## 2.5 Electro-kinetic energy conversion efficiency $(\eta_e)$

Electrokinetic energy conversion efficiency refers to the effectiveness with which the hydraulic energy is converted to electrical energy. Electrical power is extracted by the conversion of the hydraulic energy of the pressure driven flow by means of streaming potential and streaming current. The corresponding expression for energy conversion efficiency (EKEC) is given as [18]

$$\eta_e = \frac{P_{elec}}{P_{hyd}} \tag{11}$$

Here, $P_{hyd} = \Delta P \cdot Q_{in}$ is the input power per unit length of the channel due to externally imposed pressure gradient; $Q_{in}$ is the volume flow rate considering purely pressure driven flow given as $Q_{in} = w\int_{y=0}^{H} u_p dy$ and $u_p$ denotes purely pressure driven flow velocity. Harnessed electrical power is calculated following the expression $P_{elec} = \frac{I_{stream}}{2}\frac{(\Delta\psi)}{2}$. Dimensionless streaming potential takes the form $\Delta\bar{\psi}$ and dimensionless streaming current can be expressed as:

$$\bar{I}_{stream} = \int_{\bar{y}=0}^{1} (\bar{n}_+ - \bar{n}_-)\bar{u}\,d\bar{y} \tag{12}$$

Thus the resulting expression of $\eta_e$ becomes:

$$\eta_e = \frac{P_{elec}}{P_{hyd}} = \frac{P_{elec}}{P_{elec} + \Delta p \int_{y=0}^{H} u\,dy} \tag{13}$$

where $u$ denotes the flow velocity due to coupled effect of imposed pressure gradient and reverse electro-osmosis. In terms of non dimensional parameters, EKEC finally takes the form:

$$\eta_e = \frac{\Delta\bar{\psi}\left[\int_0^1 (\bar{n}_+ - \bar{n}_-)\bar{u}\,d\bar{y}\right]}{\Delta\bar{\psi}\left[\int_0^1 (\bar{n}_+ - \bar{n}_-)\bar{u}\,d\bar{y}\right] + \int_0^1 \bar{u}\,d\bar{y}} \tag{14}$$

## 3. RESULTS AND DISCUSSIONS

The present problem is described by a set of coupled governing differential equations $(1-4)$. These governing equations in their dimensionless form $(8-a,b,c,d)$ are solved in conjunction with the appropriate set of boundary conditions (section 2.3) to obtain the resulting flow field, charge and ionic species distributions. The nonlinear nature of the differential equations doesn't allow us to obtain an analytical solution and hence we resort to the numerical solution of the same with the help of commercial finite element package of COMSOL Multiphysics. The numerical simulations are performed based on dimensionless controlling parameters, chosen carefully considering the practical values of different physical parameters involved in the present study. Bulk ionic concentration is varied in between $10^{-4}$ mM and $10^{-2}$ mM while the induced surface charge density is varied in between $10^{-4}$ and $10^{-2}$ C/m$^2$.[12,14] Local permittivity variation is considered in present work in order to incorporate high surface charge density effect. Hydrodynamic slip length is considered to lie in between ~ $10-10^2$ nm [28,29,31]. The above-mentioned values result in a nondimensional Debye length $(\bar{\lambda})$ of the order of ~ $O(10^{-2})$ to $O(10^{-1})$ and normalized surface charge density $(\bar{\sigma})$ of the order of ~ $O(10^1)$ to $O(10^2)$ for a channel height taken in the order of micron $H \sim O(10)\,\mu m$. Thus the normalized hydrodynamic slip length varies of the order of ~ $O(10^{-3})$ to $O(10^{-2})$. Whether the consideration of ionic advection is justified or not is governed by a non dimensional number called ionic Péclet number $(Pe)$ which is defined as: $Pe = u_{ref} H / D = u_e H / D$. In the present study, $u_r$ is assumed $(u_r = u_e / u_{ref})$ to be 1, hence, $u_{ref} = u_e$. Typical values of the dimensional parameters involved in the definition of $(Pe)$ are as follows: $D \sim O(10^{-9})\,m^2/s$, $H \sim O(10)\,\mu m$ and $u_e \sim O(10^{-5})$ m/s where $u_e = \left(\varepsilon_0 k_b^2 T^2 / z^2 e^2 \mu H\right)$ and μ is taken as 0.001 Pa.s. Based on these values, ionic Péclet number has been considered of the order of ~ $O(10^{-1})$ to $O(10^0)$. The steric-factor $(\nu)$ range is taken as ~ $0.01-0.25$ [48]. Considering the variation in surface charge density of the solid obstructions $(\sigma_p)$ with respect to that of the channel wall $(\sigma_w)$, a charge density ratio is specified as $\sigma_r = \sigma_p / \sigma_w$ and varied in the range ~ 1–10. A typical value of water dipole moment is considered as $p_0 = 4.79 D$ [49], and the number density of water molecules in the bulk is taken as $n_{0,w} = 55$ M. The ionic Péclet number is chosen to be 0.1, and the parameters $\sigma_r, p$ are fixed to 1 and 0.8 respectively for the sake of comparison of results unless otherwise mentioned.

## 3.1 Validation

The grid independence study of the numerical solution presented here has been elaborated in the **Appendix.** Here, we further attempt to validate our proposed model for planar slit channel, incorporating practical considerations of steric effect and local permittivity variation and no slip boundary condition at the wall against experimental observations made previously.

*Experimental perspective*

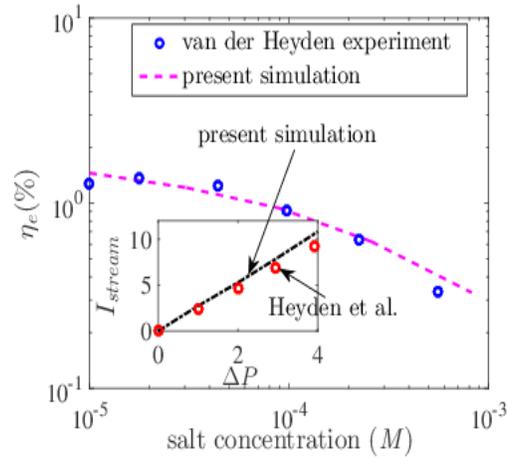

**Fig.-2** Variation of $\eta_e$ as a function of salt concentration for planar slit channel [20]. Inset depicts variation of streaming current (in pA) as a function of pressure difference, $\Delta P$ (in bar) for the same. [51]. The markers represent the experimental observation and the lines represent the present numerical simulation.

Fig.-2 (inset) shows variation of streaming current as a function of pressure for a planar slit channel of height 140 nm and 0.33M salt concentration. This result depicts that the present numerical simulation is in good agreement with the experimental observations made by Heyden et al.[51]. In a later investigation, van der Hayden and co-workers[20] pefromed an experimental investigation on power generation by pressure driven transport of ions in nanofluidic confinement. They also developed a theoretical model to predict the maximum output power and energy conversion efficiency as a function of slat concentration for a fixed channel height having different electrical boundary conditions. While comparing the presently adopted numerical methodology with the work of Hayden et al.[20] for a chemically varying surface charge density, we find that the present results are in good agreement (Fig.-2, Main-panel) with the corresponding experimental data for a channel height of 490 nm with imposed pressure difference of 4 bar.

### 3.2.1 *Effects of $R_p$ and $\sigma_r$ on $\eta_e$:*

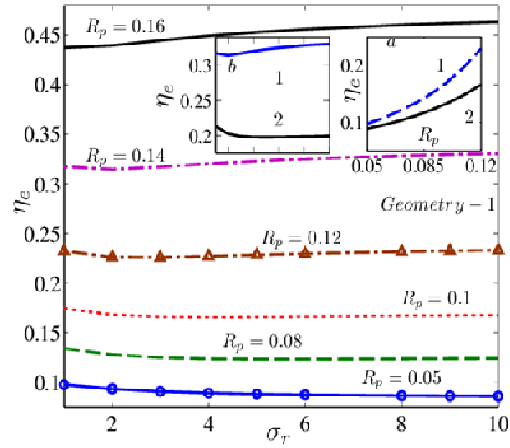

**Fig.-3** Variation of $\eta_e$ as a function of $\sigma_r$ for different obstacle radius ($R_p$). (For geometry-1 having pore-distribution in a $N \times M$ matrix form, $N = 5, M = 2$). **Insets (a)** and **(b)** depict variation of ($\eta_e$) as a function of ($R_p$) and $\sigma_r$ respectively, for two different geometric configurations.(1 and 2 stands for geometry-1 and geometry-2 respectively). Other parameters are: $\bar{\lambda} = 0.1, L_s = 0, \nu = 0, Pe = 0.1$.

Variation of $\eta_e$ as a function of $\sigma_r$ for different obstacle radius $(R_p)$ is depicted by Fig.-3. Inset-(a) portrays a gradually increasing trend of energy conversion efficiency $(\eta_e)$ with an increase in obstacle size $(R_p)$ for a fixed number of charged obstacles in a row $(N = 5)$. This trend attributes to the fact that with increase in flow obstacle size, the effective flow passage, which is in between two successive flow obstacles and channel wall, gets narrower leading to the enhancement of the streaming potential and the energy conversion efficiency. Also it is observed that, for a fixed size of flow obstruction, geometry-1 (inset-a) is more effective in terms of energy conversion perspective. The reason behind this trend may be attributed to the presence of charged cylindrical obstacles in a regular pattern causing narrower flow passages (in between two successive flow obstacles and channel wall) in comparison to zig-zag pattern. This leads to effective channel narrowing causing drastic rise in $\eta_e$ consistent with previous experimental studies. Inset-(b) depicts variation of $(\eta_e)$ with $\sigma_r$ for two different geometric arrangements. Both the insets depict that geometry-1 is more effective in terms of energy conversion perspective. Fig.-3(main panel) portrays variation of $(\eta_e)$ as a function of $\sigma_r$ which throws light towards the critical size of charged obstacle and charge density ratio to be maintained in order to achieve optimum energy conversion efficiency. It is observed that, for an obstruction size $R_p <= 0.1$ (i.e 10% of the channel height), $(\eta_e)$ decreases initially with an increase in $\sigma_r$ (up to $\sigma_r \sim 2$) and remains almost invariant with further increase in $\sigma_r$. For obstacle size above $R_p = 0.1$, $(\eta_e)$ decreases initially with an increase in $\sigma_r$ (up to $\sigma_r \sim 2$), then shows a gradually increasing trend with further increase in $\sigma_r$ up to a value of $\sigma_r \sim 9$, and saturates thereafter. That increment is observed to be more significant for larger size of flow obstacles $(R_p = 0.16)$. Hence, it will be more effective in terms of energy conversion perspective to operate at $\sigma_r = 1$ for a pore size $R_p <= 0.1$ and $\sigma_r >= 9$ for $R_p > 0.1$ in order to achieve optimum energy conversion efficiency.

### 3.2.2 *Effect of non-dimensional pitch length $(p)$ on $\eta_e$:*

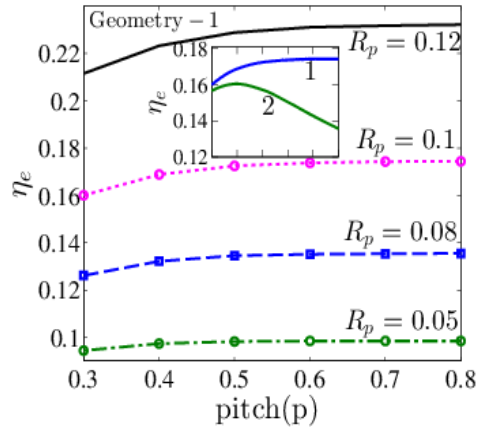

**Fig.-4** Variation of $\eta_e$ as a function of normalized pitch length $(p)$ for different size of flow obstructions for Geometry-1(a $N \times M$ matrix arrangement, $N = 5, M = 2$). Other parameters are: $\sigma_r = 1, Pe = 0.1, L_s = 0, v = 0, \bar{\lambda} = 0.1$. **Inset** depicts the same variation for two different geometric arrangements.(1 and 2 stands for geometry-1 and geometry-2 respectively)

Effect of variation of $R_p$ on $(\eta_e)$ is portrayed in Fig.-4 (for geometry-1) as a function of non dimensional pitch length $(p)$. In this context, it is noteworthy to mention that, normalized pitch length $(p) = 0.5$ signifies that the pitch length is 5 times the normalized Debye length $\bar{\lambda}$. For a value of $R_p >= 0.05$ (i.e 5% of channel height), a slight increase in $\eta_e$ is observed initially with an increase in $p$ (up to $p \sim 0.6$) and then remains invariant with further increase in $p$. This increment of $\eta_e$ (for $p <= 0.6$) is more significant for higher values of $R_p$. This sets a critical operating value of normalized pitch length for a fixed obstacle size in perspective of optimal energy conversion efficiency. Inset depicts the same variation for two different geometric arrangements. (For a fixed $N$ and $R_p$). It is observed that for geometry-2, $\eta_e$ increase with $p$ up to $p \sim 0.4$ and then decreases gradually with further increase in pitch length which sets the critical operating value of pitch length.

### 3.3 *Effect of variation of $\bar{\sigma}$ on $\Delta\bar{\psi}$ and $\eta_e$:*

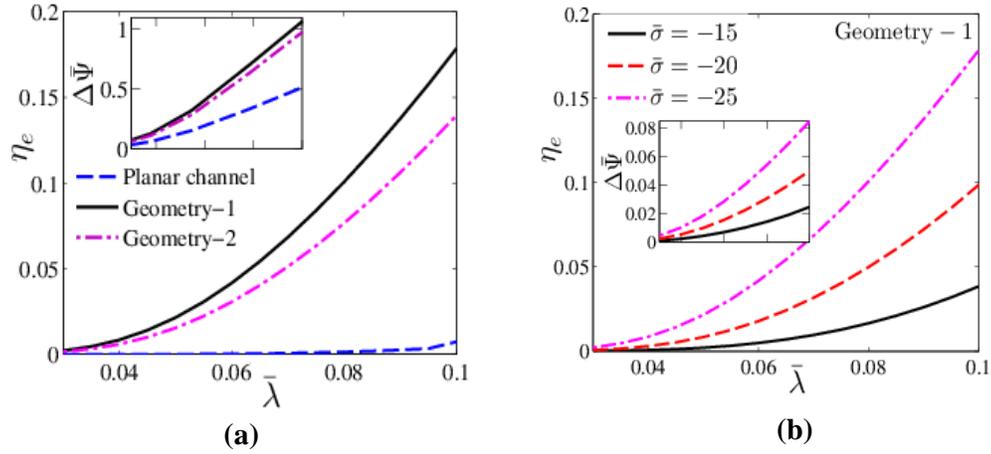

(a)                (b)

**Fig.-5(a)** Variation of $\eta_e$ as a function of $\bar{\lambda}$ for 3 different geometric patterns.(Variation of $\Delta\bar{\psi}$ with $\bar{\lambda}$ is shown in **5.(a)-Inset**). **(b) (Inset +main panel)** $\Delta\bar{\psi}$ and $\eta_e$ variation as a function of $\bar{\lambda}$ for 3 different values of dimensionless surface charge density $\bar{\sigma}$ (-15,-20,-25) for geometry-1.(geometry-1 consists of flow obstacles arranged in a $N \times M$ matrix arrangement, $N = 5, M = 2$ ). Other parameters are $\sigma_r = 1, L_s = 0, v = 0, Pe = 0.1, R_p = 0.1$.

Variation of $\eta_e$ as a function of $\bar{\lambda}$ is depicted in Figure.-5(a) for 3 different geometric arrangements where the inset shows the corresponding variation for normalized streaming potential $(\Delta\bar{\psi})$. The other parameters are set to the following values: $\sigma_r = 1, L_s = 0, v = 0, Pe = 0.1, R_p = 0.1$. From the above mentioned figure it is observed that, geometry-1 is the most effective one in perspective of conversion efficiency consistent with previous observations. Both streaming potential and $\eta_e$ further shows a gradually increasing trend with an increase in normalized Debye length $(\bar{\lambda})$ as also observed previously[12,21,52]. Figure-5(b), inset and main panel portrays variation of $\Delta\bar{\psi}$ and $\eta_e$ respectively as a function of $\bar{\lambda}$ for 3 different values of dimensionless surface charge density $\bar{\sigma}$ (-15,-20,-25). From the above mentioned figure it is observed that highest $\eta_e$ is obtained for the largest value of $\bar{\sigma}$. The reason behind this trend may be explained as follows: with increase in surface charge density, bulk ionic concentration increase causing reduction in the characteristic EDL thickness and enhancement in both positive and negative ionic species number density in the electrolyte solution. For higher value of surface charge density, this leads to simultaneous increment of the absolute magnitude of both streaming and conduction current. But the increment in streaming current occurs at much faster rate due to its coupling with flow field causing significant enhancement in streaming potential (which is the ratio of streaming current to conduction current) and conversion efficiency. As surface charge density decrease, the corresponding streaming current to conduction current ratio gradually decreases causing a decrement in streaming potential, streaming current and conversion efficiency[12,15,17,53]. Fig.-5(b) depicts that an energy conversion efficiency up to~18% can be achieved corresponding to a $\bar{\lambda} \sim 0.1$ and $\bar{\sigma} = -25$, (for geometry-1) which is almost 4.5 times of that obtained from planar-channel ($\eta_e \sim 4\%$) under similar condition.

### 3.4 *Effect of* **variation** *of* $Pe, L_s$ *on* $\Delta\bar{\psi}$ *and* $\eta_e$:

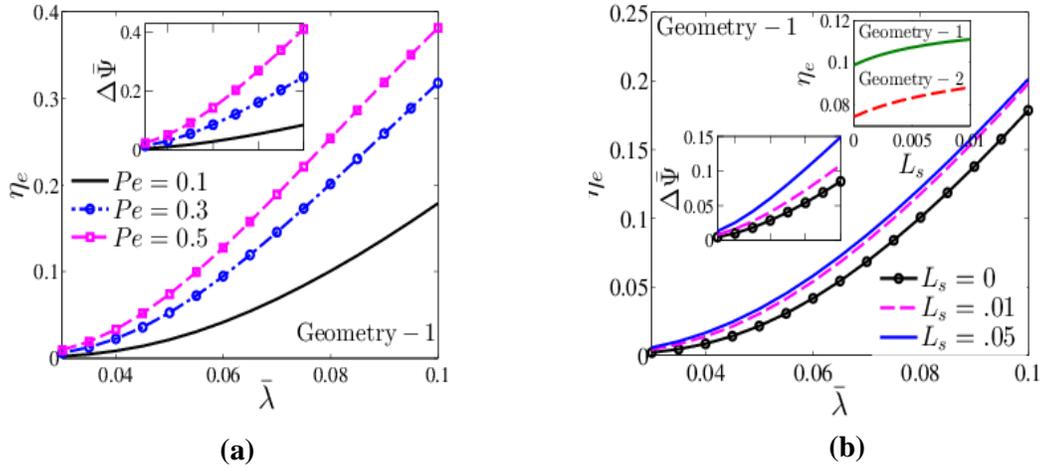

(a)            (b)

**Fig.-6(a,b)** shows variation of $\eta_e$ as a function of $\bar{\lambda}$ for two non dimensional parameters namely ionic Péclet number ($Pe$) and normalized hydrodynamic slip length ($L_s$) respectively for geometry-1.( geometry-1 consists of flow obstacles arranged in a $N \times M$ matrix arrangement, $N = 5, M = 2$ ).Insets of both the figures depict variation of $\Delta\bar{\psi}$ as a function of $\bar{\lambda}$ for the two above mentioned non dimensional parameters. Other parameters are $\sigma_r = 1, \nu = 0, R_p = 0.1$.

Variation of $\eta_e$ as a function of $\bar{\lambda}$ for 3 different values of ionic Péclet numbers ($Pe$ =0.1, 0.3, 0.5) is observed in Fig.-6(a) where the inset shows the corresponding results for $\Delta\bar{\psi}$ variation. For a fixed value of normalized Debye length $(\bar{\lambda})$, slopes of the curves are observed to be much steeper with increase in ionic Péclet number for both the cases. Such an increasing trend of streaming potential and energy conversion efficiency is observed due to the combined interplay between enhanced convective transport of ionic species and increase in EDL thickness.

Fig.-6(b) depicts the effect of consideration of hydrodynamic slip velocity at channel wall and the surface of solid obstacle on the variation of $\eta_e$ as a function of normalized Debye length. The inset shows the same on the variation of normalized streaming potential $\Delta\bar{\psi}$. It is observed that, consideration of hydrodynamic slip leads to the enhancement of both streaming potential and conversion efficiency as compared to its no-slip counterpart. The reason behind such an increasing trend is that for a fixed value of $\bar{\lambda}$, increase in normalized hydrodynamic slip length $(L_s)$ causes an increase in the average flow velocity and effective flow rate of the electrolyte solution through the channel which leads to enhancement of streaming potential, current and conversion efficiency.[13,15,17]

### 3.5 Fabrication perspective

Most of the significant works reported by different research groups[14,19,20] on electrokinectic energy conversion have been accomplished either on planar slit channel or capillaries having channel size in order of nanometers to the best of our knowledge. However, the energy conversion efficiency reported by them is significantly low [12,14,19,20]. Besides, nano channels are more difficult to fabricate due to high precession, fabrication cost, time and skill comparing to the micro channels. In the present study we have shown that, by introducing solid cylindrical flow obstacles in a micro channel (channel size ~ O(10) micron) leads to a significant enhancement of energy conversion efficiency comparing to that obtained from unrestricted micro-channels. Nevertheless, it must be commented that the reported method can also be effective in terms of fabrication perspective and cost. Towards fabrication of such flow restricted channels, any well-established protocol can be followed such as performing photolithography followed by soft lithography process for the channel fabrication purpose[54]. Photolithography has to be performed on silicon wafer for the preparation of master mould. The photo mask fabrication technique, essential for the photolithography process, plays a pivotal role in the implementation of solid cylindrical flow obstacle in form of structured solid pores. Mask should be fabricated in such a way that only the flow-able regions of the rectangular micro-channel (area of the flow obstructions subtracted from the rectangular channel area) remains exposed to the UV light source. Exposing the UV light through the photo mask over the spin-coated negative photoresist wafer followed by etching results in the master mould formation having protrusion of the rectangular channel with hollow cylindrical holes inside (height of the hole is same as the channel height). This is followed by soft lithography performed using PDMS. Finally, the microchannel structure on PDMS has to be peeled off and bonded over a glass or silicon wafer to obtain the desired geometry.

**Conclusion:**

Here, we have investigated the effectiveness of introducing charged cylindrical flow obstruction in planar slit type channel on effective augmentation of streaming potential and energy harvesting efficiency considering combined effects of finite ionic size, local permittivity variation, and wall hydrodynamic slip. In particular, we pinpoint the effect of flow obstacle size, centre to centre distance between two consecutive obstacles (pitch length) and the ratio of surface charge density of the pore to channel wall and pore arrangement (regular/zig-zag) on energy conversion efficiency. The important observations from the present study are enumerated as follows:

1. Regular geometric pattern (geometry-1) turns out to be more effective in perspective of energy conversion between two types of geometric patterns employed in the present study.

2. Introducing flow obstruction leads to a significant enhancement of energy conversion efficiency(~18% with geometry-1 and ~13% for geometry-2) with respect to planar channel.(efficiency~4%) for $\bar{\sigma} = -25, L_s = 0, \nu = 0, R_p = 0.1$.

3. Energy conversion efficiency is found to increase with increase in size of flow obstacle.

4. It is observed that, implementing structured flow obstacles of radii greater than 10% of channel height and charge density ratio greater than or equal to 9 renders significant enhancement of power generation efficiency. On the other hand, it is effective to operate at a charge density ratio ~1 for obstacles radii less than or equal to 10% of channel height.

5. An effective normalized pitch length of magnitude 0.6 or above has to be maintained in order to obtain optimum energy conversion efficiency.

6. Energy conversion efficiency is found to enhance with increase in normalized surface charge density and convective transport of ions.( manifested by ionic Peclet number)

7. Introducing wall hydrodynamic slip effect, pertinent to flows in narrow confinement, leads to further enhancement in electrokinetic energy conversion efficiency.

**APPENDIX:**

*Grid independence test:*

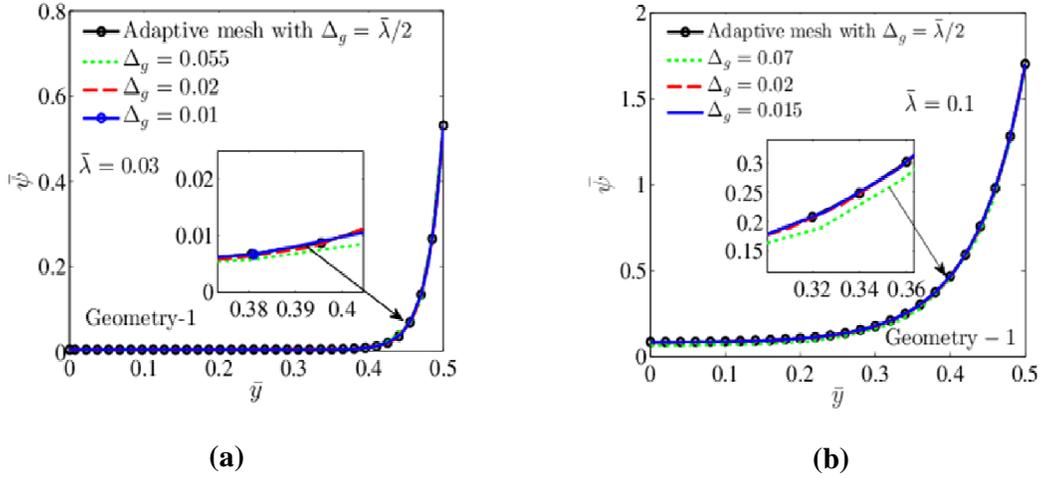

(a) (b)

**Fig.-(a),(b)** Variation of $\bar{\psi}$ as a function of $\bar{y}$ for two extreme values of $(\bar{\lambda})$ while the parameters $L_s, v$ are set to zero.(for a $N \times M$ matrix arrangement with $N=8, M=2$, Geometry-1).

The accuracy of the numerical solution has been checked with a grid independence test. Minimum grid size considered in the present simulation is lower than the normalized characteristic EDL thickness $(\bar{\lambda})$. We have chosen normalized grid/element size $\Delta_g \sim (\bar{\lambda}/2)$ throughout the domain along with adaptive mesh refinement and studied for two extreme values of $\bar{\lambda} = 0.03$ and $0.1$. Fig.-(a,b) shows variation of $\bar{\psi}$ as a function of $\bar{y}$ for 3 different values of $\Delta_g$ (0.07,0.02,0.015 for $\bar{\lambda} = 0.1$ and 0.07,0.02,0.01 for $\bar{\lambda} = 0.03$ respectively).We observe a very small deviation in the result for an element size less than or equals to 0.02 for both of the cases which eventually indicates that we have converged to a solution which is independent of the grid size. Hence all the simulations are performed taking $\Delta_g = 0.01$ (approximate number of elements are ~10000) in the present study.